\documentclass[letterpaper,titlepage,12pt]{article}
\usepackage{amssymb,amsmath,amsfonts,enumerate}
\usepackage{graphicx}

\def\clock{{\count0=\time
           \divide\count0 60
           \ifnum\count0<10 0\fi\the\count0
           \multiply\count0 -60 \advance\count0 \time
           :\ifnum\count0<10 0\fi \the\count0
         }}
\newcommand{\timestamp}{{\small\vbox{\hbox{\tt\jobname.tex}
\hbox{\the\day/\the\month/\the\year, \clock}}}}

\newcommand{\beq}{\begin{equation}}
\newcommand{\eeq}{\end{equation}}
\newcommand{\ben}{\begin{displaymath}}
\newcommand{\een}{\end{displaymath}}
\newcommand{\beqa}{\begin{eqnarray}}
\newcommand{\eeqa}{\end{eqnarray}}
\newcommand{\bea}{\begin{eqnarray}}
\newcommand{\eea}{\end{eqnarray}}
\newcommand{\bean}{\begin{eqnarray*}}
\newcommand{\eean}{\end{eqnarray*}}
\newcommand{\ba}{\begin{array}}
\newcommand{\ea}{\end{array}}
\newcommand{\bi}{\begin{itemize}}
\newcommand{\ei}{\end{itemize}}
\newcommand{\nn}{\nonumber}
\newcommand{\ie}{{\it i.e.,\,}}
\newcommand{\eg}{{\it e.g.,\,}}

\newcommand{\lp}{\left(}
\newcommand{\rp}{\right)}

\numberwithin{equation}{section}

\begin{document}

\begin{titlepage}
\begin{flushright}
\end{flushright}
\vskip 2.cm
\begin{center}
{\bf\LARGE{Self-similar critical geometries}}
\vskip 0.12cm
{\bf\LARGE{at horizon intersections and mergers}} 
\vskip 1.5cm
{\bf Roberto Emparan$^{a,b}$, Nidal Haddad$^{a}$
}
\vskip 0.5cm
\medskip
\textit{$^{a}$Departament de F{\'\i}sica Fonamental and}\\
\textit{Institut de
Ci\`encies del Cosmos, Universitat de
Barcelona, }\\
\textit{Mart\'{\i} i Franqu\`es 1, E-08028 Barcelona, Spain}\\
\smallskip
\textit{$^{b}$Instituci\'o Catalana de Recerca i Estudis
Avan\c cats (ICREA)}\\
\textit{Passeig Llu\'{\i}s Companys 23, E-08010 Barcelona, Spain}\\

\vskip .2 in
\texttt{ emparan@ub.edu, nidal@ffn.ub.es}

\end{center}

\vskip 0.3in

\baselineskip 16pt
\date{}

\begin{center} {\bf Abstract} \end{center} 

\vskip 0.2cm 

We study topology-changing transitions in the space of
higher-dimensional black hole solutions. Kol has proposed that these are
conifold-type transitions controlled by self-similar double-cone
geometries. We present an exact example of this phenomenon in the
intersection between a black hole horizon and a cosmological deSitter
horizon in $D\geq 6$. We also describe local models for the critical
geometries that control many transitions in the phase space of
higher-dimensional black holes, such as the pinch-down of a
topologically spherical black hole to a black ring or to a black
$p$-sphere, or the merger between black holes and black
rings in black Saturns or di-rings in $D\geq 6$.

\noindent 
\end{titlepage} \vfill\eject

\setcounter{equation}{0}

\pagestyle{empty}
\small
\normalsize
\pagestyle{plain}
\setcounter{page}{1}

\newpage

\section{Introduction}

Black holes in higher dimensions exhibit a pattern of phases much more
intricate than the simple situation that uniqueness theorems impose in
four dimensions. Unraveling the structure of this phase space is a
problem in which, despite the steady progress in recent years, there
remain important open issues. In particular, the topology-changing
transitions in the space of solutions at which horizons split or merge
involve all the non-linearity of Einstein's theory and lie far from the
regimes where analytic perturbative techniques of the type developed in
\cite{Emparan:2009cs,Emparan:2009at} would apply. Numerical methods,
while very valuable, face the problem that the geometries that
effect the change in topology involve curvature singularities.

\begin{figure}\begin{center}
\includegraphics[scale=.8]{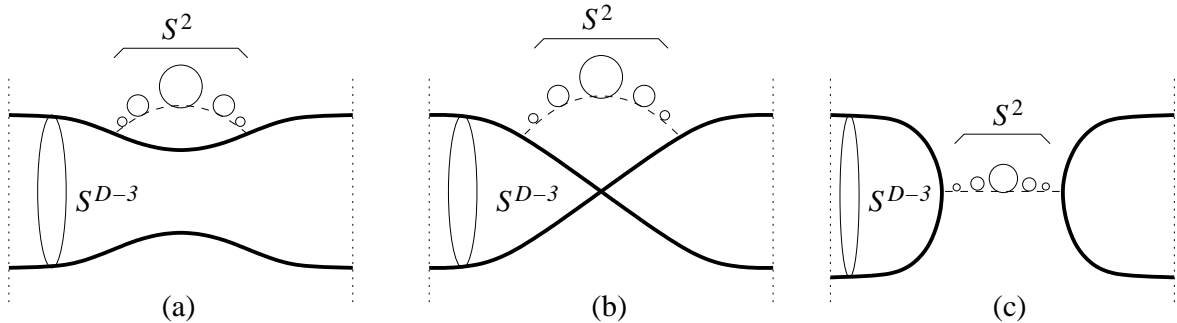} 
\end{center}
\caption{\small 
Black hole/black string transition in a Kaluza-Klein circle (the circle
runs along the horizontal axis), following \cite{Kol:2003a}. The circle fibered
over the dashed segments is the Euclidean time circle, which shrinks to
zero at the horizon: this fiber bundle describes a $S^2$. We also mark a
cycle $S^{D-3}$ on the horizon. In the black string phase (a) the $S^2$
is contractible, the $S^{D-3}$ is not, while in the black hole phase (c)
the $S^{D-3}$ is contractible, the $S^2$ is not. The transition between
the two phases is of conifold type, with the critical geometry (b)
becoming, near the pinch-off point, a cone over $S^2\times S^{D-3}$.
}\label{fig:cone1}
\end{figure}

The most studied example of this phenomenon\footnote{In this paper we
are concerned only with evolution in the phase space of stationary black
hole solutions, not with time evolution in dynamical mergers of
horizons.} is the black hole-black string transition in Kaluza-Klein
compactified spacetimes
\cite{Kol:2004ww}. The uniform string
phase branches into a non-uniform black string phase
\cite{Gubser:2001ac} whose
non-uniformity grows until a cycle along the horizon of the string
pinches down to zero size. The same pinched-off phase can be approached
from black holes localized in the KK circle, where the size of the black
hole grows until the two opposite poles of the horizon come into contact
with each other along the circle
\cite{Harmark:2002tr} 
(see ref.~\cite{Headrick:2009pv} for the state of the art in
solutions on both sides of the transition). Early in these
studies, ref.~\cite{Kol:2003a}
emphasized that the pinched-off solution plays the role of a critical
point in phase space and argued that some of its properties, in
particular the local geometry near the singular pinch-off region, should
be determined by general symmetry considerations. Specifically, it was
proposed, using an argument that we summarize in fig.~\ref{fig:cone1},
that this geometry (after Wick-rotation to Euclidean time) is locally
modelled by a self-similar cone over $S^2\times S^{D-3}$. The transition
between phases is analogous to the `conifold transition', where the
critical cone geometry can be smoothed in two ways, each one leading to
one of the phases at each side of the transition. 

Ref.~\cite{Kol:2003a} could present in exact form only the local conical
geometry asymptotically close to the pinch-off point. The details of
how, and even whether, this conical region extends to a full critical
solution in the KK circle remained open. Moreover, the non-singular
geometries that approach the critical solution are only known
perturbatively or numerically (see \cite{Kol:2003b,Asnin:2006ip} and
other works cited above). The aim of this paper is to, first, present an
exact example of a horizon-merger transition that provides strong
analytic evidence for the conifold-type picture, and second, to argue
for the universality of this picture among wide classes of
topology-changing transitions involving higher-dimensional black hole
phases. 

To this end, in section~\ref{sec:MPdSmerger} we present a complete,
exact family of geometries that approach a merger of horizons, and we
give explicitly the critical solution at the merger. The system
describes the meeting of the horizon of a black hole with a cosmological
deSitter horizon in any $D\geq 6$, with the black hole sufficiently
distorted away from spherical symmetry that it intersects the deSitter
horizon along a circle, and not on all points on the horizon. This study
confirms that the critical solution near the pinch-off is locally a
self-similar cone with the geometry anticipated in \cite{Kol:2003a}.
Section~\ref{sec:merged} is an attempt to find the geometry after the
horizons have merged into a single one. Unfortunately, the result is
only partially successful since the solution presents some pathologies.

In section~\ref{sec:othercrits} we describe the critical geometries at
the merger point in other important instances --- here we do not have a
description of the approach to the merger transition, but we can always
identify the local model for the critical geometries.
Refs.~\cite{Emparan:2007wm,Emparan:2010sx} (following
\cite{Emparan:2003sy}) have proposed the following picture for a
specific class of topology-changing transitions: a black ring of horizon
topology $S^1\times S^{D-3}$ in $D\geq 6$ becomes, as its spin
decreases, fat enough that its hole closes up. Coming from the other
side of the transition, the same phase is reached when a pinched
rotating black hole pinches off to a singularity at its axis of
rotation. In this paper we describe the local conical geometry that
controls this transition. We also extend the analysis to similar
transitions that involve black holes with horizon topology $S^p\times
S^{D-p-2}$, with $p\geq 1$ \cite{Emparan:2009cs,Emparan:2009vd}. We find
the critical conical geometries that appear when the round $S^p$ closes
off and the critical solution connects to a black hole of spherical
topology. Other conical geometries, for instance those that appear at
the merger of a black ring and a black hole, or two black rings, are
easily obtained too.

\begin{figure}
\centerline{
\includegraphics[scale=.5]{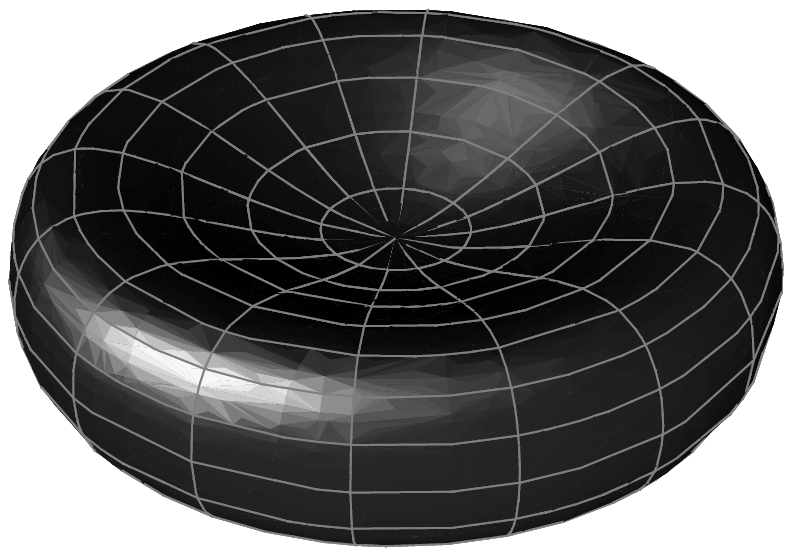}
\includegraphics[scale=.5]{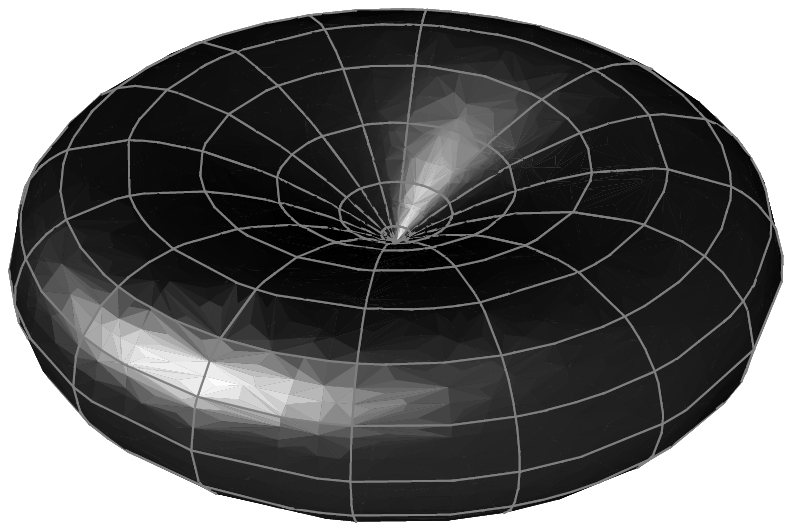}
\includegraphics[scale=.5]{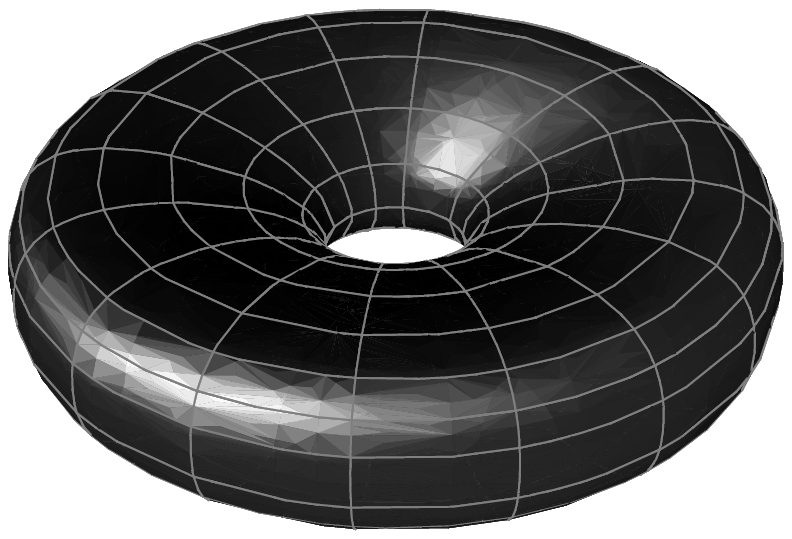}
}
\caption{\small
Black ring pinch in $D\geq 6$. The pictures are only illustrative of
expected solutions that are yet to be constructed. In
sec.~\ref{sec:othercrits} we describe the critical geometry
near the self-similar pinch-off point. 
}\label{fig:3rings}
\end{figure}
\begin{figure}
\centerline{
\includegraphics[scale=.7]{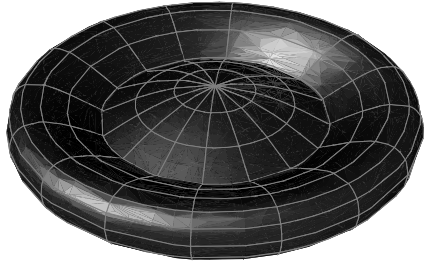}
}
\caption{\small
Circular pinch in the transition involving a black Saturn in $D\geq 6$.
}\label{fig:Sring}
\end{figure}

An interesting, perhaps unexpected, consequence of our analysis is that
the critical conical geometries appear not only when horizons merge, but
more generally when they just intersect. To understand the distinction,
note that if there is an actual transition in which two separate
horizons merge to form one horizon, the surface gravities (\ie\
temperatures) of the two horizons must approach the same value at the
critical solution. However, we find exact critical solutions in which
the two horizons have different surface gravities. These horizons can
approach and touch each other locally, \ie intersect, over a
singularity, but they cannot merge to form a single, connected horizon
over which the temperature must be uniform. Such intersection geometries
correspond to endpoints of trajectories in the space of
solutions, and not to topology-changing transitions. We find that they
also take the form of self-similar cones, but their base is not a
homogeneous space (a direct product of round spheres) but an
inhomogeneous one (a warped product).

Caveat emptor, evolution in the space of black hole
solutions, as studied in this paper, occurs along geometries that can be
quite different than in dynamical evolution in time, as the
example of the black string/black hole transition shows \cite{Lehner:2010pn}.
Presumably this is also the case in the instances we discuss here.

\section{Intersection of black hole and deSitter horizons}
\label{sec:MPdSmerger}

The so-called $D$-dimensional Kerr-deSitter solution for a rotating black hole in
deSitter space with a single rotation, as found in \cite{Hawking:1998kw} but
with a shift $\phi\to\phi+a t/L^2$, is
\beqa\label{MPdS}
ds^2&=&-\frac{\Delta_r}{\rho^2\Xi^2}\lp\Delta_\theta dt-a\sin^2\theta d\phi\rp^2+
\rho^2\lp\frac{dr^2}{\Delta_r}+\frac{d\theta^2}{\Delta_\theta}\rp\nonumber\\
&&+\frac{\Delta_\theta\sin^2\theta}{\rho^2\Xi^2}\lp (r^2+a^2)d\phi-a \lp 1-\frac{r^2}{L^2}\rp dt\rp^2
+r^2\cos^2\theta d\Omega^2_{(D-4)}
\eeqa
with
\beqa
\rho^2&=&r^2+a^2\cos^2\theta\label{defrho}\,,\\
\Delta_r&=&(r^2+a^2)\lp 1-\frac{r^2}{L^2}\rp-\frac{2M}{r^{D-5}}\,,\label{defDelr}\\
\Delta_\theta&=&1+\frac{a^2}{L^2}\cos^2\theta\,,\label{defDelth}\\
\Xi&=&1+\frac{a^2}{L^2}\,,\label{defXi}
\eeqa
and
\beq
0\leq\theta\leq \frac{\pi}{2}\,,\qquad 0\leq \phi\leq 2\pi\,.
\eeq
The metric satisfies $R_{\mu\nu}=(D-1)L^{-2}g_{\mu\nu}$. When $M=0$ this
is deSitter spacetime in
`ellipsoidal coordinates'. When $M$ is non-zero and positive, there is a
range of values of $M$ and $a$ for which the function $\Delta_r$ has two
real positive roots that correspond to the black hole and cosmological
horizons. With non-zero rotation $a\neq 0$, both horizons are distorted
away from spherical symmetry. For reasons that will become apparent,
we only consider $D\geq 6$.

\subsection{Horizon intersection}

\begin{figure}\begin{center}
\includegraphics[scale=.5]{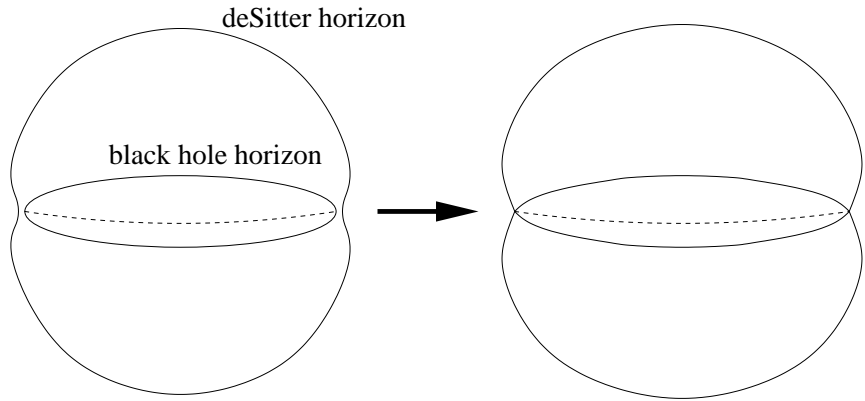} 
\end{center}
\caption{\small 
Sketch of black hole-deSitter horizons in the approach to the solution in
which the black hole touches the deSitter (cosmological) horizon along its
equator. Only
in $D\geq 6$ does the black hole admit a large enough distortion away
from spherical symmetry to allow this type of configuration.
}\label{fig:bhdS}
\end{figure}

We want to take a limit in which the black hole grows and its
horizon touches the cosmological horizon, in such a
way that this occurs not uniformly over all of the horizon, but only along the
`equator' at $\theta=\pi/2$, as illustrated in
fig.~\ref{fig:bhdS}. 
There is an intuitive reason why this should be possible in $D\geq 6$:
in these dimensions, Myers-Perry black holes rotating along one plane
can become very flat and thin, effectively like disks of a black
membrane \cite{Emparan:2003sy}. The blackfold approach of
\cite{Emparan:2009cs,Emparan:2009at} allows to construct a
\textit{static} configuration in which this disk extends along a plane
and touches the deSitter horizon. This blackfold
construction is easy to perform and gives an approximate solution for
the intersecting-horizon configuration. However, we will not give its
details since we can obtain the complete exact solution that it is an
approximation to. 

The appropriate limit of \eqref{MPdS} is 
\beq\label{lim1}
a,M\to\infty
\eeq
keeping fixed
\beq\label{lim2}
\mu=\frac{2M}{a^2}\,.
\eeq
Even if we are taking the rotation parameter $a$ to be very large,
this does not mean that the black hole rotates very rapidly relative to
the cosmological horizon. By taking $M$ large we are also making the
black hole size grow. The relative drag between the two horizons
increases and as a consequence they approach corotation, with the
relative angular velocity decreasing. Eventually, when the black hole
touches the deSitter horizon, the configuration becomes manifestly
static with our choice of coordinates. However, as we shall see
presently, since the distortion away from spherical symmetry remains in
the limit, the horizons touch only along the equator. 

Taking the above limit in \eqref{MPdS}, we find
\beqa\label{MPdScone}
ds^2&=&L^2(d\theta^2+\sin^2\theta d\phi^2)\nonumber\\
&&+
\cos^2\theta\lp -\lp 1-\frac{\mu}{r^{D-5}}-\frac{r^2}{L^2}\rp dt^2
+\frac{dr^2}{1-\frac{\mu}{r^{D-5}}-\frac{r^2}{L^2}}+r^2 d\Omega^2_{(D-4)}\rp\nonumber\\
&=&L^2\lp d\theta^2 +\sin^2\theta d\phi^2 \rp
+\cos^2\theta\; ds^2(\textrm{Schw-dS}_{D-2})
\,,
\eeqa
where Schw-dS$_{D-2}$ denotes the Schwarzschild-deSitter geometry in
$D-2$ dimensions.
When $\mu=0$ this factor becomes $(D-2)$-dimensional
deSitter (dS$_{D-2}$) and the whole
geometry is dS$_D$ spacetime of radius $L$. 
The singularity at $\theta=\pi/2$ in this
metric is just a coordinate artifact. The cosmological horizon at $r=L$ has the usual
geometry of a round $S^{D-2}$.

Our actual interest is in the solutions with $\mu\neq 0$. Then the singularity at
$\theta=\pi/2$ is a true one where the curvature diverges.
Setting $\theta-\pi/2=z/L$, then near $z=0$ the geometry asymptotically becomes
\beq\label{coneline}
ds^2\to dz^2+L^2d\phi^2 +\frac{z^2}{L^2}ds^2(\textrm{Schw-dS}_{D-2})\,.
\eeq
This has the form of a cone over Schw-dS$_{D-2}$, spread along the circle
generated by $\phi$.

Before we took the limit, the solution \eqref{MPdS} in the parameter
range of interest described two separate horizons, black hole and
cosmological. In the limiting solution, these are still present. The
equation
\beq
\frac{r^2}{L^2}+\frac{\mu}{r^{D-5}}=1
\eeq
with
\beq
0<\frac{\mu}{L^{D-5}}<\frac{2}{D-5}\lp\frac{D-5}{D-3}\rp^{\frac{D-3}{2}}
\eeq
has two real positive roots for $r$, for the black hole
and cosmological horizons in the Schw-dS$_{D-2}$ sub-spacetime. These
are the limits of the black hole and cosmological horizons of the
original solution \eqref{MPdS} in $D$ dimensions. In \eqref{MPdScone}
these two horizons come to touch each other
along the circle $\theta=\pi/2$.

The continuously self-similar structure of a cone around this intersection
point is
apparent in \eqref{coneline}. If we analytically continue to Euclidean
time $\tau$, then the $(\tau,r)$ part of the Schw-dS$_{D-2}$ geometry
describes a two-sphere, generically with a conical defect (of
codimension 1) at either the cosmological or black hole horizons since
their temperatures are not the same for
generic values of $\mu/L^{D-5}$. Except for this defect the
Euclidean geometry of Schw-dS$_{D-2}$ is a warped product of
$S^2_{(\tau,r)}$ and $S^{D-4}$. So the Euclidean continuation of
\eqref{coneline} is a cone over this warped $S^2\times S^{D-4}$, times
the $\phi$ circle. This cone is essentially the local model for the critical
geometry that \cite{Kol:2003a} had proposed.

Note, however, that what we have found is a more general cone than
in \cite{Kol:2003a}. In \eqref{MPdScone} the
temperatures of the black hole and of the cosmological horizons need not
be the same. For instance, in
configurations where the black hole has the shape of a very thin
pancake, which occur when $\mu\ll L^{D-5}$, the black
hole temperature is clearly much higher than the temperature of the
cosmological horizon.\footnote{The blackfold method reproduces these
configurations to leading order in $\mu/L^{D-5}\ll 1$. In the exact
solutions one can take a limit to focus on the region very close to the
axis $\theta=0$ and recover the geometry of a black 2-brane of thickness
much smaller than $L$.}

When their temperatures are different, the two horizons can approach
each other and \textit{intersect}, but not evolve beyond the
intersection to \textit{merge} and form a single, connected horizon.
Since the temperature over the latter must be uniform, a merger requires
that the temperatures of the two horizons approach the same value at the
intersection.
Ref.~\cite{Kol:2003a} focused on merger transitions, but we have seen that the
appearance of self-similar conical geometries is a more general feature
of intersections of horizons, even when they cannot merge. For
the Lorentzian solutions the difference in temperatures does not imply
any pathology, so it is natural to consider
these configurations as well.

The parameter $\mu$ in the critical geometry \eqref{MPdScone} can be
adjusted so that the two horizons have the same surface gravity. In fact
it is possible to consider not only this solution, but an entire
subfamily of solutions of \eqref{MPdS} in which the two separate
horizons, black hole and cosmological, have the same surface gravity
even away from the critical merger geometry. This subfamily of solutions
describes a rotating black hole with the same temperature as the
cosmological horizon, \ie a `rotating Nariai' solution, and is of some
interest in itself, so we describe it next.

\subsection{Isothermal solutions}\label{subsec:isotherm}

The limit of the solutions
\eqref{MPdS} where the two separate horizons have equal temperatures, or
surface gravities, is
defined in appendix~\ref{app:nariai}. The mass and the radial coordinate
are fixed to values $M=M_0$ and $r=r_0$ that depend on the
rotation parameter $a$.
Using for simplicity units where $L=1$, the metric that results is
\beqa\label{nariaidS}
ds^2&=&C\rho_0^2(-\sin^2\chi d\tilde t^2+d\chi^2)
+\frac{\rho_0^2}{\Delta_\theta}d\theta^2\nonumber\\
&&+\frac{\Delta_\theta\sin^2\theta}{\rho_0^2\Xi^2}
\left((r_0^2+a^2)d\tilde\phi-2 C r_0 a\Xi\cos\chi
d\tilde t\right)^2
+r_0^2\cos^2\theta d\Omega^2_{(D-4)}\,.
\eeqa
where
\beq
\rho_0^2=r_0^2+a^2\cos^2\theta\,,
\eeq
and $\Delta_\theta$, $\Xi$ as in \eqref{defDelth}, \eqref{defXi} (with $L=1$) and 
\beq
0\leq\chi\leq\pi\,,\qquad 0\leq\theta\leq \frac{\pi}2\,,
\qquad 0\leq \tilde\phi\leq 2\pi\,.
\eeq
The constants $r_0$ and $C$ are determined in terms of $a$ by
\beq\label{barabarr}
a^2=r_0^2\frac{D-3-(D-1)r_0^2}{(D-3)r_0^2-(D-5)}
\eeq
and
\beq\label{bigC}
C=\frac{(D-3)r_0^2-(D-5)}{(D-1)(D-3)r_0^4-2(D-5)(D-1)r_0^2+(D-5)(D-3)}\,.
\eeq
Actually, we could take the only parameter in the solution to be
$r_0$ instead of $a$, since they are related in one-to-one
manner in the range of interest, which is 
\beq
0\leq a\leq \infty\,,\qquad \frac{D-5}{D-3}\leq r_0^2\leq \frac{D-3}{D-1}\,
\eeq
(note this implies $r_0^2<1$).

In going from \eqref{MPdS} to \eqref{nariaidS}, the cohomogeneity
of the geometry has been reduced from two to one, the only non-trivial
dependence being now on $\theta$. It is manifest that the
surface gravities at the two horizons, at $\chi=0$ and $\chi=\pi$, are
equal. However, there is 
a relative angular velocity between them,
\beq
\Omega_{rel}=
\frac{4Cr_0 a\Xi}{r_0^2+a^2}\,.
\eeq
For a merger of the two horizons to be possible, this relative motion
between them must disappear at the critical solution.

In the conventional static limit $a\to 0$, where $r_0^2\to(D-3)/(D-1)$,
eq.~\eqref{nariaidS} becomes the Nariai limit of
Schw-dS$_{D}$, namely
the direct product geometry
\beq
ds^2\stackrel{a\to 0}{\longrightarrow} \frac{1}{D-1}(-\sin^2\chi d\tilde t^2+d\chi^2)+
\frac{D-3}{D-1}d\Omega_{(D-2)}^2
\eeq
which Wick-rotates to $S^2\times S^{D-2}$. In this solution, and in all
the solutions with finite $a$, the two horizons remain separate.

The limit that we are interested in, where the two horizons touch at the
equator $\theta=\pi/2$, lies at $a\to\infty$, with
$r_0^2\to(D-5)/(D-3)$. Then $\Omega_{rel}\to 0$
(as we had already observed in the previous section) and
\beq\label{nariaicone}
ds^2\stackrel{a\to \infty}{\longrightarrow} 
\frac{1}{D-3}\cos^2\theta(-\sin^2\chi d\tilde t^2+d\chi^2)+
d\theta^2+\sin^2\theta d\tilde\phi^2+\frac{D-5}{D-3}\cos^2\theta
d\Omega_{(D-2)}^2\,.
\eeq
This is indeed the same geometry that results if we take the Nariai limit
of the Schw-dS$_{D-2}$ geometry inside \eqref{MPdScone}. We have
obtained it here along a particular one-parameter subfamily of
the solutions \eqref{MPdS}.

In the region close to the intersection of horizons, with small
$\theta-\pi/2=z/L$, the geometry \eqref{nariaicone} becomes, after
rotation to Euclidean time (and restoring $L$),
\beq\label{spreadcone}
ds^2\to dz^2+L^2 d\tilde\phi^2 +\frac{z^2}{D-3}\lp d\Omega^2_{(2)}+ (D-5) d\Omega^2_{(D-4)}\rp\,,
\eeq 
which is exactly the kind of double-cone geometry predicted by
the arguments
in \cite{Kol:2003a}.

\bigskip

We have studied the solutions in which the initial black hole rotates
along a single plane, but it is straightforward to extend this to the
general solutions with rotation in an arbitrary number of planes and
then obtain intersections along odd-spheres instead of circles.
Appendix~\ref{app:generalbhds} explains this construction.

\section{Merged solution}\label{sec:merged}

It is natural now to look for an exact solution for the
geometry after the two horizons have merged into a single one. Here we
describe our attempt at finding this solution. 
In contrast to the pre-merger solution of the previous section, the
merged solution that we find is not entirely satisfactory --- its
Lorentzian section is complex, and it has a naked singularity. It is
unclear to us whether this is a deficiency somehow intrinsic to the (cosmological)
set up that we are considering, or whether there is another solution for
the merged configuration. Readers not interested in the details can
safely jump to section~\ref{sec:othercrits}.

One might expect, by analyticity in the space of solutions, that the
solution after the black hole and cosmological horizons have merged
should be in the same family as \eqref{nariaidS}, but in a different
parameter range than before the merger --- \ie in the merged solution
the parameter $a$ extends beyond the range we have been considering in
sec.~\ref{subsec:isotherm}. Actually, we will consider not only
\eqref{nariaidS}, or the initial Kerr-dS solution \eqref{MPdS}, but the
Kerr-NUT-dS family. This is the largest known family of solutions that
appear appropriate for this task. As in section~\ref{sec:MPdSmerger},
away from the critical solution we consider the Lorentzian section of
the geometries.

It is useful to first identify the main properties that the solution
must possess. First, the family of solutions should obviously have a limit to the
critical geometry \eqref{nariaicone}.

Second, from the general picture of the conifold-type
transition, in the merged solution there should be a `Lorentzian
two-sphere' (like the one that $\tilde t$ and $\chi$ describe in
\eqref{nariaicone}) that is contractible to zero, and a $S^{D-4}$ that
is not. Observe that this, and the first
condition, rule out the simple possibility that when the two horizons merge
we recover dS$_D$.

\begin{figure}\begin{center}
\includegraphics[scale=.95]{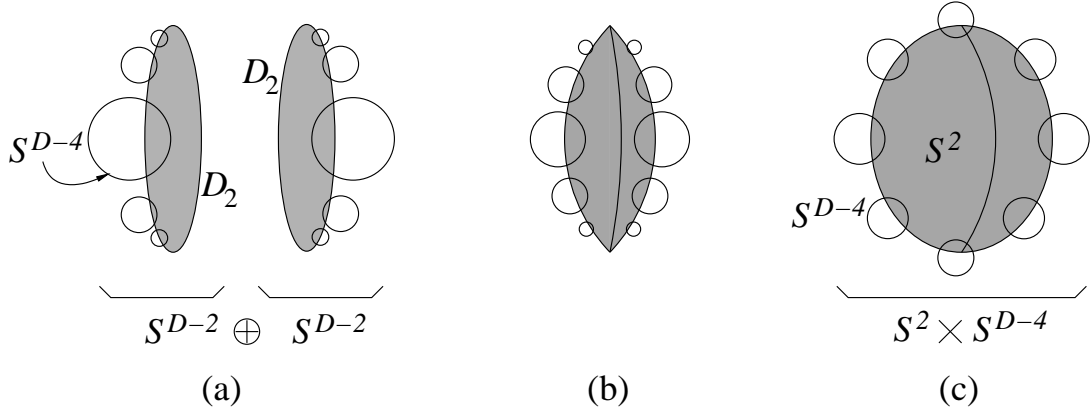} 
\end{center}
\caption{\small Horizon geometry in (a) the black hole-deSitter phase;
(b) critical solution; (c) merged solution.
}\label{fig:MPdS}
\end{figure}
Finally, we can easily determine the topology of the spatial
sections of the horizon. As illustrated in
figure \ref{fig:MPdS}, before the merger the horizon is the sum of two
$S^{D-2}$. Each of these
spheres can be viewed as the result of fibering the (topological) disk
$D_2$, parametrized by $(\theta,\phi)$, with spheres $S^{D-4}$ whose
size goes to zero at the boundary of the disk at $\theta=\pi/2$. In the
critical geometry, the two disks meet at their edges: they form a
topological $S^2$. The spheres $S^{D-4}$ fiber over this $S^2$, with
their sizes shrinking to zero at the circle where the two disks meet:
this is the horizon of the critical solution. When the two horizons
merge to form a single one, the $S^{D- 4}$ do not shrink to zero
anywhere, so the horizon of the merged solution has the
topology $S^2\times S^{D-4}$.\footnote{Away from the critical solution,
this could be a non-trivial bundle. This will not be important in our
analysis.}

Consider now the Kerr-NUT-dS solution in $D$-dimensions \cite{Klemm:1998kd}, in
units of $L=1$,
\beqa\label{KNdS}
ds^2&=&\rho^2\lp\frac{dr^2}{\Delta_r}+\frac{du^2}{\Delta_u}\rp
-\frac{\Delta_r}{\rho^2\Xi^2}
\lp(1+a^2 u^2)dt-a(1-u^2)d\phi\rp^2\nn\\
&&+\frac{\Delta_u}{\rho^2\Xi^2}
\lp (r^2+a^2)d\phi-a(1-r^2)dt\rp^2
+r^2u^2d\Omega^2_{(D-4)}
\eeqa
with $\Delta_r$ and $\Xi$ as in \eqref{defDelr}, \eqref{defXi}, and
\beq
\rho^2=r^2 + a^2 u^2\,,\qquad 
\Delta_u=(1-u^2)(1+a^2 u^2)+\frac{2N}{u^{D-5}}\,.
\eeq
Our notation and choice of parametrization is different than in \cite{Klemm:1998kd}, but more
adequate for our purposes.
If we set the NUT parameter\footnote{In the four-dimensional solution,
the usual NUT parameter is not exactly the same as $N$ here. The
relation is nevertheless easily found.} $N=0$ and redefine
$u=\cos\theta$ we recover
\eqref{MPdS}. Note that having $N\neq 0$ in $D\geq 6$ prevents $u$ from
reaching zero, since the last term in $\Delta_u$ would blow up.

We are interested in solutions with a single temperature, so we take the
same limit as in appendix~\ref{app:nariai} to obtain a generalization of
\eqref{nariaidS},
\beqa\label{nariaiNdS}
ds^2&=&C\rho_0^2(-\sin^2\chi d\tilde t^2+d\chi^2)
+\frac{\rho_0^2}{\Delta_u}du^2\nonumber\\
&&+\frac{\Delta_u}{\rho_0^2\Xi^2}
\left((r_0^2+a^2)d\tilde\phi-2 C r_0 a\Xi\cos\chi
d\tilde t\right)^2
+r_0^2u^2 d\Omega^2_{(D-4)}\,,
\eeqa
where
\beq
\rho_0^2=r_0^2+a^2u^2\,,
\eeq
and $r_0$, $C$ are the same functions of $a^2$ as in
sec.~\ref{subsec:isotherm}. The
only non-trivial dependence of the metric is on $u$.

There are two parameters, $a$ and $N$. In sec.~\ref{subsec:isotherm} we set
$N=0$ and $0\leq a^2\leq \infty$, with the critical phase being reached
as $a\to\infty$. Here we
extend the solutions into a new parameter range by
considering
negative values of $a^2$, specifically
\beq
-\infty\leq a^2< -1
\eeq
as covered when $r_0$ varies in\footnote{This range of $a^2$ is also
obtained with $1<r_0^2\leq\infty$, but one can see this is not adequate
for obeying the required behavior.}
\beq
\frac{D-5}{D-1}< r_0^2\leq \frac{D-5}{D-3}\,.
\eeq
Note that this implies $0<r_0^2<1$, and also $M_0 <0$. 
It is convenient to define
\beq
\alpha^2=-a^2\,,\qquad
\hat C=Ca^2\,,
\eeq
so that $\alpha^2,\hat C>0$, and
\beqa
\Sigma(u)&=&\frac{\rho_0^2}{a^2}=u^2-\frac{r_0^2}{\alpha^2}\,,\\
\Upsilon(u)&=&\frac{\Delta_u}{a^2}
=(1-u^2)(u^2-\alpha^{-2})+\frac{2\hat N}{u^{D-5}}\,,
\eeqa
where we have conveniently absorbed a power of $\alpha$ in $\hat N$.
The metric reads\footnote{Refs.~\cite{Lu:2004ya,Mann:2005ra} give the Euclidean version of this
solution in a different parametrization that is more elegant but less
appropriate for our purposes.}
\beqa
ds^2&=&\hat C\Sigma(-\sin^2\chi d\tilde t^2+d\chi^2)
+\frac{\Sigma}{\Upsilon}du^2\nonumber\\
&&+\frac{\Upsilon}{\Sigma\Xi^2}
\left((\alpha^2-r_0^2)d\tilde\phi+2i\hat C
r_0\alpha \Xi\cos\chi d\tilde t\right)^2
+r_0^2u^2 d\Omega^2_{(D-4)}\,,
\eeqa
This is a complex metric, with imaginary $g_{\tilde t\tilde\phi}$,
since we are taking the rotation parameter $a$ into imaginary values. We
will return to this issue below.

We take $u$ to vary in an interval for which
$\Sigma$ and $\Upsilon$ are both non-negative. 
The zeroes of $\Upsilon$, which limit this interval, depend on the NUT
parameter $\hat N$, which so far has been free. When $\hat N=0$, 
the range in which $\Upsilon$ is positive is
\beq
\alpha^{-1}\leq u\leq 1\qquad (\hat N=0)\,
\eeq
(we need only consider $u>0$).
Since for $\alpha<\infty$, $u$ is never zero, we fulfill one of the topological requirements
on the merged solution: the $S^{D-4}$ never shrinks to zero size.
However, since $\alpha^{-1}>r_0/\alpha$, we have that $\Sigma(u)$,
although positive, is never zero and therefore the
Lorentzian-$S^2_{\tilde t, \chi}$ is not contractible. This can be
remedied by turning on the parameter $\hat N$ and tuning it so that the
smallest of the two relevant roots of $\Upsilon$ moves to
the value $r_0/\alpha$; the other will be $u_1>1$. Then we take
\beq
\frac{r_0}{\alpha}\leq u \leq u_1\,.
\eeq
Note that since now $\Sigma$ and $\Upsilon$ both have a simple zero at $u=r_0/
\alpha$, the $\tilde\phi$ circle has finite size there.

While it is easy to solve for the required value of $\hat N$ as a
function of $\alpha$, we are
mostly interested in the regime where
$\alpha$ is very large, in which we approach
the critical solution. Then one finds
\beq
\hat N\approx \frac{1}{D-3}\lp\frac{D-5}{D-3}\rp^{\frac{D-5}2}\frac1{\alpha^{D-3}}\,,
\qquad 
\frac{r_0}{\alpha}\approx \sqrt\frac{D-5}{D-3}\frac1{\alpha}\,,
\qquad
u_1\approx 1+\hat N\,.
\eeq
Note that $\hat N\to 0$ as $\alpha\to\infty$.

With these parameter choices, we obtain a solution in which 
\begin{itemize}
\item
as $\alpha\to \infty$ the metric becomes that of the
critical solution \eqref{nariaicone} (with $u=\cos\theta$), 
\item there is a
contractible Lorentzian-$S^2$ and a non-contractible $S^{D-4}$.
\item the constant-$\tilde t$ sections of the horizon, 
where $\sin\chi=0$, have topology $S^2\times
S^{D-4}$. The $S^2$ is made of the two disks at $\chi=0,\pi$, joined
along $u=r_0/\alpha$.
\end{itemize}
Therefore, this one-parameter family of solutions satisfies
all the properties that we required at the beginning of
the section.

Unfortunately, these geometries have two significant
shortcomings. First, they are complex, and then it is unclear
whether it is sensible to talk about a horizon. Second,
and perhaps worse, when the Lorentzian-$S^2_{\tilde t, \chi}$ shrinks to
zero at $u=r_0/\alpha$, it does not do so smoothly. Near this point,
the $(u,\tilde t,\chi)$ part of the geometry behaves (up to constant
factors) as
\beq
(u-r_0/\alpha)(-\sin^2\chi d\tilde t^2+d\chi^2)+du^2
\eeq
which is singular at $u=r_0/\alpha$.\footnote{The fact that these solutions
have $M<0$ might be behind this.}

The problem of the Lorentzian metric being complex can be remedied by
going to the Euclidean section. However, the trouble then comes back in
that it does not seem possible to have a regular, real Euclidean section
for the solution \textit{before} the merger. It seems we cannot have the
real transition both ways.

We have not found any other way of obeying the topology
requirements of the merged solution using the family of metrics
\eqref{KNdS} than with these parameter choices. It is unclear whether
there may be a more general solution that is better behaved.

\section{Critical geometries for other topology-changing transitions}
\label{sec:othercrits}

In sec.~\ref{sec:MPdSmerger} we have presented an exact instance of an
intersection of horizons that gives a satisfactory account of all the
aspects of the pre-merger transition conforming to the analysis of
\cite{Kol:2003a}. Thus it seems justified to look for
local models for the critical geometries in other horizon-merger transitions
that are expected to occur for higher-dimensional black holes. We will
see that there exist self-similar cone geometries with the adequate
properties for all these transitions.

\paragraph{Black ring pinch.}

The simplest new critical geometry that we describe corresponds to the
transition between a black ring with horizon topology $S^1\times
S^{D-3}$ and a black hole with horizon topology $S^{D-2}$ in $D\geq 6$. In the black
hole phase, the horizon geometry develops a pinch along the rotation axis,
which grows until it pinches off in the critical solution. Coming from
the black ring side, the ring becomes fatter until its central hole
closes up. The pinch-off occurs on the rotation axis, so we can expect
that asymptotically close to this point the rotation is
negligible. Thus, the self-similar geometry around this point will be
locally a static cone.

\begin{figure}\begin{center}
\includegraphics[scale=.9]{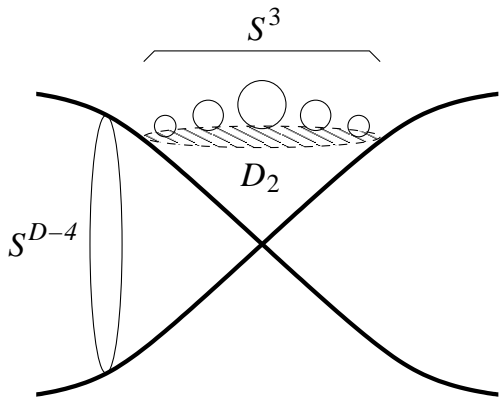} 
\end{center}
\caption{\small Critical geometry at the `black ring pinch' transition
between a black ring and a topologically spherical black hole. Relative
to fig.~\ref{fig:cone1}, the main difference is that the dashed segment
is replaced by a disk $D_2$. In general, for a
$p$-sphere pinch, the disk $D_2$ is replaced by a ball $B_{p+1}$, and
the $S^3$ by a $S^{p+2}$ ($p=0$ is the case in fig.~\ref{fig:cone1}).
}\label{fig:pinch}
\end{figure}

In the black ring phase the Euclidean time circle fibers over disks
$D_2$ that fill the ring's hole, to form a $S^3$. In addition, on the
horizon we can find spheres $S^{D-4}$ that shrink to zero size at the
inner rim of the ring. Instead, in the black hole phase these $S^{D-4}$
do not shrink anywhere in the region close to the axis, while the
previously described $S^3$ does shrink to zero there. Hence, we have an
instance of a conifold-type transition, with the critical geometry being
a cone over $S^3\times S^{D-4}$,
\beq
ds^2=dz^2+\frac{z^2}{D-2}\lp 2d\Omega^2_{(3)}+(D-5)d\Omega^2_{(D-4)}\rp
\eeq
(see fig.~\ref{fig:pinch}; this can be obtained by rotating the critical
solution of
fig.~\ref{fig:cone1} around a vertical axis.).
The Lorentzian version of the geometry is
\beq
ds^2=dz^2+\frac{2z^2}{D-2}\lp -\cos^2\chi dt^2+d\chi^2+\sin^2\chi d\phi^2
+\frac{D-5}{2}d\Omega^2_{(D-4)}\rp
\eeq
with the horizon being at $\chi=\pi/2$. 

Note that this conifold transition can not occur for five-dimensional
black rings since the required cone does not exist.

\paragraph{$p$-sphere pinch.}

Refs.~\cite{Emparan:2009cs,Emparan:2009vd} generalized black rings with
horizon $S^1\times S^{D-3}$ to solutions with horizon $S^p\times
S^{D-p-2}$, with odd $p$, where the $S^p$ is a contractible cycle --- we
refer to it as an `odd-sphere blackfold'. Like in a black ring, the
$S^p$ has to rotate in order to maintain its equilibrium at a given
radius. The approximate method used in
\cite{Emparan:2009cs,Emparan:2009vd} is only valid as long as
the blackfold is thin, \ie the $S^p$ is much larger than the
$S^{D-p-2}$, but it is natural to expect that when the angular momentum
decreases, the sphere $S^p$ shrinks until its inner hole closes up, like
in the example of the black ring. At that point the configuration makes
a transition to a topologically-spherical black hole.

In this case the inner hole is a ball $B_{p+1}$ over which the Euclidean
time circle fibers, shrinking to zero at the boundary of the ball: this
is a $S^{p+2}$. The size of this sphere remains finite in the
odd-sphere-blackfold phase, and shrinks to zero at the axis in the
topologically-spherical black hole phase. On the other hand the horizon has a
$S^{D-p-3}$ that shrinks to zero size in the odd-sphere
blackfold phase, and remains finite (near the axis) in the black hole
phase. Thus we have another conifold transition, this time with a critical geometry 
\beq
ds^2=dz^2+\frac{p+1}{D-2}z^2\lp d\Omega^2_{(p+2)}+\frac{D-p-4}{p+1}d\Omega^2_{(D-p-
3)}\rp\,,
\eeq
and in the Lorentzian version
\beq
ds^2=dz^2+\frac{p+1}{D-2}z^2\lp -\cos^2\chi dt^2+d\chi^2
+\sin^2\chi d\Omega^2_{(p)}
+\frac{D-p-4}{p+1}d\Omega^2_{(D-p-3)}\rp
\eeq
with $0\leq\chi\leq\pi/2$ and horizon at $\chi=\pi/2$.

Note that the boundary of $B_{p+1}$ is connected when $p>0$ so, unlike
in sec.~\ref{sec:MPdSmerger} (which was locally the case $p=0$ smeared
along a circle), in these cases the merger necessarily involves only one
horizon.

In the odd-sphere blackfold, when the angular momenta are
not all equal, the $S^p$ is not
geometrically round. One may reduce some of the angular momenta while
the others are kept fixed. In this case we expect a transition to an
ultraspinning topologically spherical black hole, controlled by a local
geometry of the type above with a collapsing sphere of lower dimensionality.

\paragraph{Circular pinch.}

Finally, we describe another type of critical pinch-off geometry that is
expected to occur in the phase space of higher-dimensional rotating
black holes. This is the case of a circular pinch at finite radius on
the plane of rotation. This mediates the transition between, for
instance, a black Saturn in $D\geq 6$ and a topologically spherical
black hole with a circular pinch. The same local critical geometry also
appears when two black rings in $D\geq 6$ merge.

The two black objects that merge are rotating, but as they approach we
can go to a reference frame that is asymptotically corotating
with the two horizons, and in this frame the geometry near the pinch-off
point looks again static. This is then locally like in
fig.~\ref{fig:cone1}, spread over the direction of the circle
where the horizons touch.  The local model for the critical
geometry is the same as \eqref{spreadcone}. Note again that this cannot occur in
five dimensions. 

It is actually possible to consider also intersecting geometries in
which the two black objects have different temperatures, as
in \eqref{coneline}, where the horizons
touch on a cone but cannot merge. What does not seem possible is to have the two
horizons touch over a cone if they have relative non-zero velocity.
Presumably, an attempt to force two such horizons to touch gives a
stronger singularity.

Given these examples, it is straightforward to extend them to cones for
many other transitions that are expected to occur for higher-dimensional
black holes.

\section{Concluding remarks}

The main features of the phase space of neutral, asymptotically flat,
higher-dimensional black holes
are controlled by solutions in three different regions:
\begin{enumerate}
\item[(i)] Large angular momenta.
\item[(ii)] Bifurcations in phase space.
\item[(iii)] Topology-changing transitions.
\end{enumerate}
The regime (i) is captured by the blackfold effective theory
\cite{Emparan:2009cs,Emparan:2009at}. Regions (ii) are controlled by
zero-mode perturbations of black holes that give rise to bifurcations
into new families of solutions. The initial conjectures about these
points \cite{Emparan:2003sy,Emparan:2007wm} have been confirmed and
extended in \cite{Dias:2009iu}. In this paper we have begun to
explore regions (iii) in $D\geq 6$ and provided local models for the
critical geometries that effect the topology change.

We have presented an example where we can study in a detailed exact
manner the geometry in a topology-changing transition, at least in one
of the sides of the transition and at the critical point itself. The
critical geometry conforms precisely to the predictions of
\cite{Kol:2003a}. It seems valuable to have an exact and simple analytic
model of one such transition from which one can extract further details.
In particular, a more detailed study of how the conical geometries,
including the large class of examples in appendix~\ref{app:generalbhds},
are resolved away from the critical point is probably of interest.

We have also seen that self-similar cone geometries occur when two
horizons intersect but cannot merge since their temperatures are
unequal. In this case the cone is over a warped product, not a direct
one, and the arguments of \cite{Kol:2003a} would not apply.
Nevertheless, the extension we have found is a natural one: the
direct-product Nariai solution at the base of the cone is simply
replaced by the more general Schwarzschild-deSitter solution. This
allows to study intersecting horizons in more generality.

Topology-changing transitions among five-dimensional asymptotically flat
rotating black holes do not fall within the class studied in this paper.
Indeed, the phase diagram of five-dimensional black holes, say with a
single spin, already reveals that the transitions are controlled by a
different class of critical geometry, not of the conifold type. The same
is true of transitions that are effectively in that same class by having
some directions smeared along the intersection (\eg black Saturn or
black di-ring mergers), and in general those that involve a collapsing
$S^1$. The study of these transitions deserves a
separate investigation.

\section*{Acknowledgments}

Work supported by MEC FPA2010-20807-C02-02, AGAUR 2009-SGR-168 and CPAN
CSD2007-00042 Consolider-Ingenio 2010. We acknowledge hospitality and a
highly stimulating environment at the Centro de Ciencias de Benasque
Pedro Pascual during the Strings Workshop and the Gravity Workshop in
July 2011, and in particular we are grateful to the participants for very
useful discussions with many of them.

\appendix

\section{Equal-temperature limit}\label{app:nariai}

In this appendix we define a limit of the Kerr-deSitter solutions
\eqref{MPdS} and the Kerr-NUT-deSitter solutions \eqref{KNdS} that
generalizes the well-known Nariai limit of the Schw-dS
solution\footnote{Ref.~\cite{Chong:2004hw} describes essentially the same limit, but
in a different parametrization.}. We use
units where the cosmological radius is $L=1$.

In order for the two horizons of \eqref{MPdS} to have equal surface
gravities, or equal temperatures, the value of the radial coordinate $r$
must be the same for both, \ie $\Delta_r$ must have a double root
$r=r_0$,
\beq
\Delta_r(r_0)=\partial_r\Delta_r(r_0)=0\,.
\eeq
These two conditions determine the values of $M$ and $a$ as functions of
$r_0$,
\beq
M_0=r_0^{D-3}\frac{(1-r_0^2)^2}{(D-3)r_0^2-(D-5)}\,,\qquad
a^2=r_0^2\frac{D-3-(D-1)r_0^2}{(D-3)r_0^2-(D-5)}\,.
\eeq
Although their radial coordinates coincide, the proper radial distance between
the two horizons goes to a finite non-zero limit. We can pry open the
space between them by
first introducing a small parameter $\varepsilon$ that takes us slightly away
from the limit, with a new `radial' coordinate
$\chi$,
\beq
r=r_0(1-\varepsilon\cos\chi)\,,
\eeq
and
\beq
M=M_0\lp 1- \frac{r_0^2}{C (r_0^2+a^2)(1-r_0^2)}\varepsilon^2\rp\,,
\eeq
where $C$ is as in \eqref{bigC}. Then redefine appropriately the Killing
coordinates $t$ and $\phi$ to
\beq
t=C \frac{r_0^2+a^2}{r_0}\frac{\tilde t}{\varepsilon}\,,\qquad
\phi=\tilde\phi +C\frac{a(1-r_0^2)}{r_0}\frac{\tilde t}{\varepsilon}\,.
\eeq
Finally, take the limit $\varepsilon\to
0$ to find the finite metric \eqref{nariaidS}.

The limit fixes one of the two dimensionless free parameters of the
original solution, leaving $r_0$, or $a$, as the only parameter.
In this paper we are interested in approaching the solutions that
satisfy \eqref{lim1}, \eqref{lim2}. In the one-parameter family, this
corresponds to
\beq
r_0^2\to \frac{D-5}{D-3}\,.
\eeq 
Indeed in this limit
\beq
a\to\infty\,, \qquad M_0\to\infty\,, 
\eeq
while 
\beq
\frac{M_0}{a^2}\to \frac{1}{D-5}\lp\frac{D-5}{D-3}\rp^{\frac{D-3}{2}}
\eeq
remains finite.
Also, in this limit $Ca^2\to 1/(D-3)$.

\section{General black hole-deSitter intersections}\label{app:generalbhds}

Here we extend the analysis of sec.~\ref{sec:MPdSmerger} to the situation
where the black hole rotates in an arbitrary number of planes. The starting
solution is the general Kerr-deSitter metric as given in ref.~\cite{Gibbons:2004js},
whose presentation we follow closely.

In order to have a unified description for even and odd $D$, we
introduce 
\beq
\epsilon=(D-1)~\textrm{mod}~2\,.
\eeq
The spacetime dimension is then $D = 2n + \epsilon + 1$, where $n$
is the number of orthogonal rotation planes. On each of these, we choose angles
$\phi_i$ with period $2\pi$. We also introduce an overall radial
coordinate $r$ and $n + \epsilon$ direction cosines $\mu_i$ satisfying
\beq
\sum_{i=1}^{n+\epsilon} \mu_i^2 =1\,,
\eeq
with $0\leq \mu_i\leq 1$, $i=1,\dots,n$. In even $D$ the
$\mu_{n+1}$ has range $-1\leq \mu_{n+1}\leq 1$.

The solution is characterized by a mass parameter $M$ and $n$ rotation
parameters $a_i$, and its metric is
\beqa
ds^2 &=& - W\lp 1 -\frac{r^2}{L^2}\rp dt^2
 +\frac{2M}{U}\lp W dt
 -\sum_{i=1}^n \frac{a_i\, \mu_i^2}
  {1+\frac{a_i^2}{L^2}}\, d\phi_i\rp^2
 + \sum_{i=1}^n \frac{r^2 + a_i^2}{1 + \frac{a_i^2}{L^2}}\,\mu_i^2\,
    d\phi_i^2 \nn\\
&&+ 
\frac{U}{V-2M}dr^2 + 
\sum_{i=1}^{n+\epsilon} \frac{r^2 + a_i^2}{1 + \frac{a_i^2}{L^2}}d\mu_i^2
 + \frac{1}{W(L^2- r^2)}\lp \sum_{i=1}^{n+\epsilon} 
\frac{\lp r^2 + a_i^2\rp \mu_i\, d\mu_i}{1 + \frac{a_i^2}{L^2}}\rp^2 \,,
\eeqa
where
\beq 
U = r^{\epsilon}\, \sum_{i=1}^{n+\epsilon} \frac{\mu_i^2}{r^2 + a_i^2}\, 
\prod_{l=1}^n (r^2 + a_l^2)\,,
\eeq
\beq
V= r^{\epsilon-2}\, \lp 1-\frac{r^2}{L^2}\rp \, 
   \prod_{i=1}^n (r^2 + a_i^2)\,,
\eeq
and
\beq
W = \sum_{i=1}^{n+\epsilon} \frac{\mu_i^2}{1+\frac{a_i^2}{L^2}}\,.
\eeq

We take the limit in which we send a
number $s$ of rotation
parameters to infinity, all at the same rate. For now, 
the remaining $n-s$
ones are set to zero, so
\beqa
&a_j\rightarrow \infty\,,\qquad &j=1,\dots,s\,,\nn\\
&a_k= 0\,,\qquad & k=s+1,\dots,n\,,
\eeqa
with 
\beq\label{sbound}
2s\leq D-4\,.
\eeq
At the same time we send $M\to\infty$ in such a way that
\beq\label{mufin}
\mu=\frac{2M}{\prod_{j=1}^s a_j^2}
\eeq
remains finite (compare to appendix~A of \cite{Emparan:2003sy}).

We introduce an angular variable $\theta$ such that
\beq
\sum_{j=1}^{s}\mu_j^2=\sin^2\theta\,,\qquad \sum_{k=s+1}^{n+\epsilon}
\mu_k^2=\cos^2\theta\,,
\eeq
with range $0\leq\theta\leq\pi/2$.
With this variable, the metric on the unit $S^{D-2}$ is written as
\beq
\sum_{i=1}^{n+\epsilon} d\mu_i^2+\sum_{i=1}^n \mu_i^2 d\phi_i^2=d\Omega_{(D-2)}^2
=d\theta^2+\sin^2\theta\; d\Omega_{(2s-1)}^2
+\cos^2\theta\; d\Omega_{(D-2s-2)}^2\,.
\eeq
After some labor one finds the limiting geometry
\beq
ds^2=L^2\lp d\theta^2+\sin^2\theta d\Omega_{(2s-1)}^2\rp
+\cos^2\theta \lp
-f(r) dt^2
+\frac{dr^2}{f(r)}
+r^2d\Omega_{(D-2s-2)}^2\rp
\eeq
with
\beq
f(r)=1-\frac{\mu}{r^{D-2s-3}}-\frac{r^2}{L^2}\,.
\eeq
Near $\theta=\pi/2$ this has the form of a cone over Schw-dS$_{D-2s}$,
spread over a sphere $S^{2s-1}$. When $s=1$ we recover \eqref{MPdScone}.
The restriction \eqref{sbound} on the number $s$ of `ultraspins'  
guarantees that the Schw-dS$_{D-2s}$ factor in the limit geometry is at
least four-dimensional.

It is easy to generalize the limit to
\beqa
&a_j\rightarrow \infty\,,\qquad &j=1,\dots,s\,,\nn\\
&a_k~~\mathrm{finite}\,,\qquad & k=s+1,\dots,n\,,
\eeqa
again with finite $\mu$ in \eqref{mufin} to obtain the geometry
\beq
ds^2=L^2\lp d\theta^2+\sin^2\theta d\Omega_{(2s-1)}^2\rp
+\cos^2\theta\; ds^2\lp\textrm{Kerr-dS}_{(D-2s)}\rp\,,
\eeq
where Kerr-dS$_{(D-2s)}$ has the finite $a_k$ as rotation parameters. The
Euclidean version of the latter solution is known to contain many
interesting Einstein metrics, in addition to products $S^2\times
S^{D-2s-2}$, such as Page's metric for the non-trivial $S^2$ bundle over
$S^2$ \cite{Page:1979zv} and higher-dimensional generalizations thereof
(see \eg\ \cite{Lu:2004ya} and references to it). Our construction
results in cones over all these spaces. It may be interesting to
study them in more detail.

\end{document}